\documentclass[
 reprint,showkeys,showpacs,
 amsmath,amssymb,
 aps,prd,
onecolumn
]{revtex4-2}
\usepackage[dvipsnames]{xcolor}
\usepackage[T1]{fontenc}
\usepackage{graphicx}
\usepackage{natbib}
\usepackage{hyperref}
\begin{document}
\title[Triple black hole solution]{General relativistic three body problem}
\author{Eli\v{s}ka Klime\v{s}ov\'{a}}
\email{eliska.klimesova@matfyz.cuni.cz}
\affiliation{Institute of Theoretical Physics, Faculty of Mathematics and Physics, Charles University, Czech Republic}
\author{Martin \v{Z}ofka}
\email{martin.zofka@matfyz.cuni.cz}
\affiliation{Institute of Theoretical Physics, Faculty of Mathematics and Physics, Charles University, Czech Republic}
\vspace{10pt}
\begin{abstract}
We examine analytically the restricted 3-body problem in general relativity for a set of extremally charged black holes. We apply the method of Ferrell and Eardley, which consists in perturbing the Majumdar-Papapetrou solution that describes a static set of such black holes. We extend the method from a pair of black holes to a triplet. We compare the motion of a test particle and a small black hole in the field of a rotating black-hole binary and discover a repulsive self-force acting upon the third black hole.
\end{abstract}
\maketitle
\section*{Introduction}
The field of exact solutions in general relativity has taught us a great deal about our non-isotropic and non-homogeneous universe---we can now follow, e.g., the relativistic motion of stars near the supermassive black hole in the center of the Galaxy, the swirling of gravitational waves, and the cosmic flow of galaxies. Although numerical solutions can sketch in the details missing in the bigger picture, analytical solutions are essential to build our intuition. Therefore, in this paper, we will adopt an analytical approach and explore the dynamics of a system consisting of a charged black-hole triplet moving in its own gravitational and electromagnetic fields.

One of the reasons for studying this particular system is that triplets are rather abundant in the universe, as they make up roughly 5\% to 35\% of all stellar systems in the Galaxy, depending on the mass of the primary star \cite{Shariat+El-Badry+Naoz}. Furthermore, exoplanets occur in double-star systems in about 40\% of cases \cite{Matson+Howell+Horch+Everett}. Another reason is that the dynamics of multiplets are much richer than those of doublets, even in the Newtonian setting, making them interesting in their own right.

Most of the GR studies of the 3-body problem use post Newtonian approximations of various orders. There is also a post-Minkowskian approximation \cite{Ledvinka+Schafer+Bicak} that gave a closed-form Hamiltonian for an N-body system. When applying these methods, the papers focus on the existence or generalizations of special Newtonian solutions---these include, e.g., 3 point masses on a line \cite{Yamada+Asada} and the figure-8 solution \cite{Imai+Chiba+Asada,Lousto+Nakano}. Other papers explore the circular restricted 3-body problem; they study the existence, position, and stability of the Lagrange points using two heavy sources on a common circular path and a test particle moving in their field \cite{Huang+Wu,Strong+Crescimanno}.

Similarly, we also assume a central, heavy black-hole binary with fixed circular motion while focusing on the trajectory of a third, lighter companion. Disregarding its back reaction on the central binary, its effect on the spacetime---and thus on its own motion---will be included, unlike in the test-particle cases. Additionally, the method we use does not assume a weak gravitational field, as opposed to the post-Newtonian and post-Minkowskian approaches.

Working towards this goal, we follow the method of Ferrell and Eardley \cite{FeEa87,FeEa89}, which has also been applied recently to describe exact gravitational-wave signatures of near collisions of black-hole pairs \cite{CHM,FionaDaKu}. Proceeding from a static system of an arbitrary number of extremally charged black holes, one first solves the full set of combined Einstein and Maxwell equations and obtains the Majumdar-Papapetrou solution \cite{Ma,Pa}. The gravitational attraction between the sources is exactly balanced by their electrostatic repulsion, since the charges and masses of each source are the same (provided, of course, that their charges are all of equal sign), and thus they remain static indefinitely.

One then perturbs the solution by giving the individual sources small velocities, deriving the 1\textsuperscript{st}-order ($\mathcal{O}(v)$) field equations, and solving them to the 2\textsuperscript{nd} order, $\mathcal{O}(v^2)$. Surprisingly, such a complicated system of equations can be reduced to a classical-mechanics problem controlled by a Lagrangian that determines the motion of the black holes. The Lagrangian has the form of an integral over the entire space and is expressed purely in terms of the positions and velocities of the black holes, with their masses as parameters. It only involves terms combining the positions of up to four of the masses, and hence we are dealing with 4-body interactions. To obtain the corresponding Euler-Lagrange equations for the motion of the sources, one has to evaluate the Lagrangian explicitly, during which one comes across a number of mathematical difficulties, albeit the initial Lagrangian is quite simple to write down. This has only been done for two black holes so far. We aim to extend this here by studying three black holes. Since a three-body system is known to be much more complicated than a two-body system, we had to adopt some additional assumptions that are detailed below. The main advantage of this approach is that it applies even in the strong-field regime as long as the velocities of the sources are small compared to the speed of light.

The paper is structured as follows: In Section \ref{sec:Majumdar-Papapetrou_solution_and_its_perturbation} we briefly recap the Majumdar-Papapetrou solution and its slow-motion perturbation. In Section \ref{sec:Multi_black_hole_Lagrangian} we sketch how to deal with the separate terms in the three-body Lagrangian, evaluate it to yield the classical Euler-Lagrange equations describing the motion of the black holes, and study their motion. Section \ref{sec:Einstein-Maxwell_and_Geodesics} then explores the test-particle motion in the field of a black-hole binary orbiting around its common center of mass. Finally, we compare the results of Sections \ref{sec:Multi_black_hole_Lagrangian} and \ref{sec:Einstein-Maxwell_and_Geodesics}.

\section{The Majumdar-Papapetrou solution and its perturbation}\label{sec:Majumdar-Papapetrou_solution_and_its_perturbation}

A set of any number of static black holes bearing charges equal to their masses $Q_a = M_a$\footnote{Or charges equal to minus their masses $Q_a = - M_a$.} in geometrized units is described by the Majumdar-Papapetrou solution \citep{Ma,Pa} with the metric and 4-potential
 \begin{equation}
 \text{d}s^2 = -\frac{\text{d}t^2}{\psi^2}+ \psi^2 \text{d}\vec{x} \cdot \text{d} \vec{x}, \qquad
 A = \frac{1}{\psi} \text{d}t,
 \end{equation}
where the master function $\psi(\vec{x})$ is given as
 \begin{equation}
 \psi (\vec{x}) = 1 + \sum_a \frac{m_a}{|\vec{r}_a|}.
 \end{equation}
We denote $\vec{r}_a := \vec{x}-\vec{x}_a$ throughout the paper. The sum goes over all the sources indexed by $a$ and $\vec{x}_a$ is the $a$-th black hole's position.

However, we aim to describe a more realistic situation with mutual motion, since not much of the universe is really static. We let $\vec{x}_a(t)$ vary in time and assume its time derivative, the 3-velocity $\vec{v}_a$ of the source, to be small compared to the speed of light---this is the slow-motion approximation. We thus give the black holes some non-zero initial velocities. One can easily see that such a perturbation involves a lot of non-trivial physics: while in the static case, the electrostatic repulsion canceled exactly with the gravo-static interaction, now we need to take into account various electrodynamic and general-relativistic processes. This undoubtedly very complex problem is still in the full strong-field regime of gravity (though restricted to slow motions). In light of this, it is quite amazing that such a solution can be found analytically within the framework of general relativity to the lowest order in the velocity perturbation---see the seminal paper by Ferrell and Eardley \citep{FeEa89}.
The general form of the perturbed spacetime is
\begin{equation} \begin{aligned}\label{eq:prostorocas}
 \mbox{d} s^2 &= - \psi^{-2} \mbox{d}t^2 + 2 \vec{N} \cdot \mbox{d} \vec{x} \mbox{d} t + \psi^2 \mbox{d} \vec{x} \cdot \mbox{d} \vec{x}, \qquad
 A =\frac{1}{\psi} \mbox{d} t + \vec{A} \cdot \mbox{d} \vec{x},\\
\end{aligned} \end{equation}
where the perturbation is expressed using two 1\textsuperscript{st}-order ($\mathcal{O}(v)$) 3-tensorial quantities $\vec{N}$ and $\vec{A}$ on the $t=const.$ spacetime slice.
In \citep{FeEa87,FeEa89}, the authors normalize the 4-potential to zero at infinity, i.e., $A_t = -1+ 1/\psi$. However, we shall rather stick to $A_t = 1/\psi$, following Hartle and Hawking \citep{HaHa}, this minor change leaves the equations of motion unchanged.
We write the Lagrangian to the 2\textsuperscript{nd} order in these perturbative quantities and velocities $\vec{v}_a$. Looking at the system at a constant time, the gravitational interaction and electrostatic repulsion cancel exactly, no matter the positions. The only remainder will be magnetic and gravito-magnetic (frame-dragging) interactions. Let us have two slowly moving charges $q$ separated by a distance $r$. Then a simple calculation yields an estimate of the ratio of magnetic and electric forces acting between them
$$ |\vec{F}_B| = | q (\vec{v} \times \vec{B}) | \sim \frac{q^2v^2}{r^2},\quad |\vec{F}_E| \sim \frac{q^2}{r^2}, \quad \left| \frac{F_B}{F_E} \right| \sim v^2. $$
Magnetic interactions are of the order $v^2$, hence they do influence motion in our approximation, affecting charges.

What about other interactions? In the low velocity limit of extremal BH, the gravitational reaction interactions scale as $v^5$, the electromagnetism reaction forces (currents) scale as $v^3$ \citep{GiRu86}, and frame-dragging effects remain small enough as long as the velocities are small \citep{FeEa89}. The gravitational waves emitted by the slowly rotating system are of the order $v^5$ \citep{FeEa87}. In case of coalescence, they can even be of order $v^2$ \citep{FionaDaKu}, but instead we study a mutually orbiting system. So far so good, terms of the order $\mathcal{O}(v^3)$ can be neglected.

\section{Multi black hole Lagrangian}\label{sec:Multi_black_hole_Lagrangian}

We keep perturbations through up to the 2\textsuperscript{nd} order, since we are interested in the 1\textsuperscript{st}-order field equations. We start with the prescription for total exact action in general relativity and derive the perturbed action, which is to be varied to obtain the right Einstein-Maxwell equations. The total action is given by a sum of four pieces, that is gravity, Maxwell fields, action from current and from matter.

\begin{equation}
	\label{eq:dust-action}
	S = S_{\mbox{\scriptsize{gravity}}} + S_{\mbox{\scriptsize{EM-fields}}} + S_{\mbox{\scriptsize{current}}} + S_{\mbox{\scriptsize{matter}}}.	
\end{equation}
These are to be evaluated as%
\begin{equation}
	\label{eq:dust-action-rewritten}
	S = \int \mbox{d}^4x \sqrt{-g} \cdot \left( \frac{R}{16 \pi} - \frac{F^2}{16 \pi} + A_{\mu} \rho u^{\mu} - \rho \right) .
	\end{equation}
This is to be varied with respect to the fields $\vec{N}$ and $\vec{A}$ (1\textsuperscript{st}-order quantities) to obtain equations for the metric. The mass distribution $\rho$ doesn't vary freely, it is rather regulated to provide us with the matter conservation during other variations. The field perturbations are fully determined in terms of the perturbations of matter ($\rho\psi^3, \vec{v}$) by constraint equations to the order $\mathcal{O}(v^2)$ in perturbed Einstein-Maxwell equations.

We shall use the following linear combinations $ \vec{P} = \vec{A}+\psi \vec{N}, \ \vec{Q} = \psi^2\vec{N}.$ Also $\Delta \vec{K} \equiv - 4 \pi \rho \psi^3 \vec{v}$.
After varying \eqref{eq:dust-action-rewritten} one obtains the 1\textsuperscript{st}-order field equations, which are solved by \citep{FeEa89}
\begin{equation} \begin{aligned} \label{eq:P_Q_2}
 \nabla \times \vec{P} =& -3 \psi^2 \nabla \times \vec{K} -2 \nabla \chi -3 \psi^2 \nabla \xi, \quad
 \nabla \times \vec{Q} =
 -4 \psi^3\nabla \times \vec{K} -4 \psi \nabla \chi + 4 \chi \nabla \psi -4 \psi^3 \nabla \xi, \\
\end{aligned} \end{equation}
where the ``integration constants'' $\xi, \chi$ can be determined by taking the divergence and in the black hole limit contribution of these scalar functions to the effective action vanishes \citep{FeEa87}.
Now we take the black hole limit so that the smooth distribution $\rho$ of pressureless extremally charged dust turns into a sum of point-like sources
$\rho \psi^3 \rightarrow \sum_a m_a \delta^{(3)}(\vec{x}-\vec{x}_a)$,
and from the Maxwell equation (Gauss law) for Majumdar-Papapetrou solution
$\Delta \psi = - 4 \pi \sum_a m_a \delta^{(3)} (\vec{x}-\vec{x}_a)$
we obtain
 \begin{equation}
 \label{eq:BlackHoleLimitK}
 \vec{K} = \sum_a \frac{m_a}{r_a}\vec{v}_a .
 \end{equation}

Substituting \eqref{eq:P_Q_2} into the action 
\eqref{eq:dust-action-rewritten} and performing the black hole limit while keeping terms up to $O(v^2)$, one obtains that the approximate action approaches a finite limit \eqref{eq:Seff}. This effective action manages the motion of $n$ extremally charged non-rotating black holes in slow motion approximation and can be written as a sum of two pieces \citep{FeEa87,FeEa89}
 \begin{equation}
 \label{eq:Seff}
 S_{\text{eff}}=\int \mbox{d} t \left( L_{\text{free}} + L_{\text{int}} \right).
 \end{equation}
The foliation of space-time is now obviously useful, since it allows us to write $S_{\text{eff}}$ in such a simple form. A general prescription for the Lagrangian in the black-hole limit is given by a sum of a free and an interacting part, which read (the Latin indices label the black holes) \citep{FeEa87,FeEa89}
 \begin{equation} \begin{aligned} \label{eq:Lgeneral}
  L_{\text{free}} &= \frac12 \sum_{a}m_a v_a^2 - \sum_a m_a, \qquad
 L_{\text{int}} &= \frac{3}{8\pi} \int \ \psi^2 \sum_{c \neq d} \frac{m_c m_d}{(r_c r_d)^3} \left[ \frac12 |\vec{v}_c-\vec{v}_d|^2 (\vec{r}_c \cdot \vec{r}_d) - ( \vec{v}_c \times \vec{v}_d) \cdot (\vec{r}_c \times \vec{r}_d ) \right]\mbox{d}^3x,
  \end{aligned} \end{equation}
where
 $ \vec{r}_a = \vec{x}-\vec{x}_a, \; \vec{v}_a = \frac{\mbox{d}}{\mbox{d} t} \vec{x}_a, \; \psi^2= 1 + 2 \sum_c \frac{m_c}{r_c} + \sum_{c,d} \frac{m_c m_d}{r_c r_d} . $
Due to the polynomial dependence on masses $m_i$, the Lagrangian involves only up to four-body interactions. Remarkably, \eqref{eq:Lgeneral} describes a system of $N$ interacting extremally charged black holes, the motion of which is completely given by a set of 6$N$ initial conditions. In this paper, we focus on calculating the above integral analytically, which is a daunting task for anything else than mere two black holes, for which the Lagrangian reads
\begin{equation}
\label{eq:L2B}
 L_{\text{2B}} = L_{\text{free}} + L_{\text{int}}, \
 L_{\text{free}} = - M + \frac12 M \vec{V} \cdot \vec{V} + \frac12 \mu \vec{v} \cdot \vec{v}, \
 L_{\text{int}} = \frac32 \mu M \vec{v} \cdot \vec{v} \left( \frac{1}{r} + \frac{M}{r^2} + \frac{M^2 - 2 \mu M}{3 r^3} \right) ,
\end{equation}
where $M=m_1+m_2$, $\mu = m_1m_2/M$, $\vec{V}=(m_1 \vec{v}_1+m_2 \vec{v}_2)/M$, and $r = |\vec{r}_1-\vec{r}_2|$, see \citep{FeEa89}. This effectively corresponds to translation into COM coordinates. This allows us to calculate the scattering or coalescence of two black holes. Let us suppose the two black holes approach from an infinite distance with a relative velocity $v_{\infty}$ then the approximation is valid as long as their distance $r$ fulfills
$r\gtrsim v_{\infty}^2 M $.
In the following text, we rescale the impact parameter $b$, radial distance $r$, and reduced mass $\mu$ by the mass $M$, which ultimately drops out of the equations. All quantities are thus expressed in dimensionless units, implying $\mu \in (0, 1/4]$.
The critical impact parameter  that separates the scattering and coalescing trajectories then satisfies the relation
\begin{equation}
    b_{\text{crit}}^2 - \frac{2 b_{\text{crit}}^3}{3 \sqrt{3}}-2 \mu =0,
\end{equation}
which, for equal masses, reduces to $b_{\text{crit}} = \sqrt{3+ (3/2) \sqrt{3}}$. If the oncoming black hole has exactly the critical impact parameter, it will settle down to a circular orbit with $r_{\text{circ}} = 1/3(\sqrt{3} b_{\text{crit}}-3)$ which, for two equally massive holes, further reduces to
$r_{\text{circ}}=\frac{\sqrt{3} - 1}{2}  \approx 0.366 $ and, for the limit of a massive black hole and a particle, it yields $r_{\text{circ}}=1/2$.
In the following, we shall use two black holes orbiting one another at precisely this distance, which is an exact solution of the 2-body problem.

\subsection{Three-body Lagrangian}
\label{sec:L3B}

\begin{figure}[ht]
	\centering
	\includegraphics[width = 0.4\textwidth]{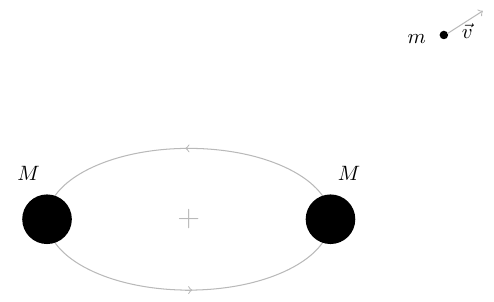}
 	\caption{A schematic image of the system with two heavy black holes orbiting around each other on a circular path and a third, lighter, companion black hole moving in their gravitational field.}
\label{fig:Sketch}
\end{figure}

Now we want to proceed similarly in order to extend the approach to three bodies.
Since the general three-body problem is known to be chaotic for most initial conditions and possesses no closed-form solution even in the Newtonian case, we imposed some assumptions to even have a chance of proceeding analytically. We assume that the triplet consists of two black holes of equal masses and a much lighter companion. The two heavy holes orbit each other on a circular path of radius $r_{21} = |\vec{x}_2 - \vec{x}_1| = r_{\text{circ}}$, given by the 2-body Lagrangian \eqref{eq:L2B}. The situation is thus analogous to the classical restricted 3-body problem, see Figure \ref{fig:Sketch}. Without loss of generality, we assume that the orbital plane of the two heavy holes is $z=0$.

Writing out the terms of the interaction Lagrangian \eqref{eq:Lgeneral} for three bodies, we split it into the 2-body Lagrangian \eqref{eq:L2B} and a remainder which is to be calculated and depends on the position and velocity of the third black hole. Several integrals appearing in the remainder cannot be evaluated analytically, so we further assume the third black hole to be far away from the black hole binary rotating around the origin: $|\vec{r}_{21}| \ll |\vec{x}_3|$. We can now decouple the Lagrangians (and evolution) of the binary and the third black hole.

\begin{figure}[ht]
	\centering
	\includegraphics[width = 14cm]{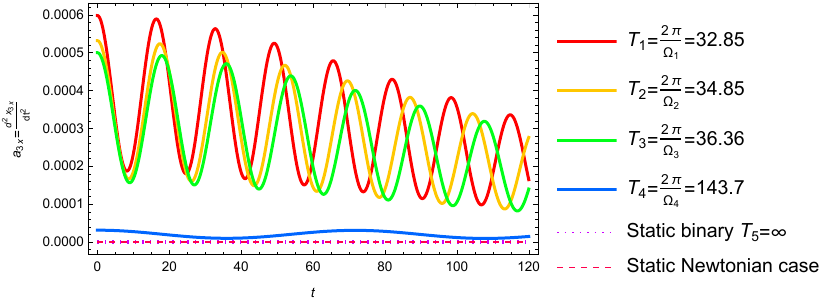}
 	\caption{Acceleration  of the third black hole dropped from rest at $\vec{x}_3(0)=(10,0,0)$ in the field of the central black hole binary orbiting with various periods $T_i$. We plot the $x$-component of the acceleration. Initially, the binary points towards the third hole. A static set of extremal black holes ($T_5=\infty$) remains static just like in Newtonian physics. Curiously and unlike in the Newtonian physics, the third black hole is always pushed out by the orbiting binary. Each peak corresponds to the closest approach of one of the central black holes. This is the coplanar case, however the results are very similar when the third black hole is dropped from a general direction. As we will see later, an initially static test particle remains static, no matter the motion of the central pair and, therefore, the curves in this plot hint at a self-force acting upon the third hole.}
\label{fig:BlackHoleAcceleration}
\end{figure}

We split the integration region spanned by $\vec{x}$ into the inner spherical region $\gamma := \{\vec{x}, |\vec{x}| \leq \Gamma \}$ of radius $\Gamma$, which is surrounded by a concentric spherical shell $\lambda : = \{\vec{x}, \Gamma < |\vec{x}| \leq \Lambda\}$ of outer radius $\Lambda$, itself surrounded by the outer concentric spherical shell $\omega : = \{\vec{x}, \Lambda < |\vec{x}| \}$. The boundaries between the regions are chosen so that $r_{21} \ll \Gamma \ll |\vec{x}_3| \ll \Lambda$. Regions $\gamma$ and $\lambda$ contribute significantly to the integral appearing in the interaction Lagrangian \eqref{eq:Lgeneral} while the outer region $\omega$ can be neglected since its contribution to the Lagrangian is proportional to $1/\Lambda$, which can be made arbitrarily small. We shall make use of the fact, that in $\gamma$ we have $|\vec{x}| \ll |\vec{x}_3|$, while in $\lambda$ we have $|\vec{x}_1|,|\vec{x}_2| \ll |\vec{x}|$, for further details, see \citep{diplomka}. Ultimately, one arrives at
\begin{equation}
\begin{aligned}
\label{eq:L3B}
 L_{3B}=-\frac32 m_3
 \Bigg (
 -\frac{|\vec{v}_3|^2}{3} + \frac23
 &+ 4 m^2 |\vec{v}_1|^2
 \left[
 \frac{1}{|\vec{x}_3|^2} + \frac{2}{r_{21} |\vec{x}_3| } + \frac{2 m }{r_{21}^2 |\vec{x}_3|}
 \right] \\
 &+ \left[
 \frac{1}{|\vec{x}_3|} + \frac{2 m }{|\vec{x}_3|^2}\right]
 2 m
 \left[ |\vec{v}_1|^2 + |\vec{v}_3|^2
 \right]
 - 4 \vec{v}_1 \cdot \vec{v}_3 \frac{ \vec{x}_3 \cdot \frac{\vec{r}_{21}}{r_{21}} }{|\vec{x}_3|^3} \frac{m^3}{r_{21}}
 \Bigg ),
\end{aligned}
\end{equation}
where $ \vec{v}_1 = - \vec{v}_2 = (- v_0 \sin (\omega t), v_0 \cos (\omega t), 0)$ and $v_0$ is the magnitude of the initial velocity of the binary, $\omega$ is their angular velocity with $v_0 = r_{21} \omega$. Also the binary black-hole masses are, for simplicity, the same $m_1 = m_2 \equiv m$. The above Lagrangian describes asymptotically free motion of the third black hole with mass $m_3 \ll m$. This can also be seen in Fig. \ref{fig:BlackHoleAcceleration}: as the third hole recedes from the binary, its acceleration gradually vanishes.

\begin{figure}[ht]
	\centering
	\includegraphics[width = 14cm]{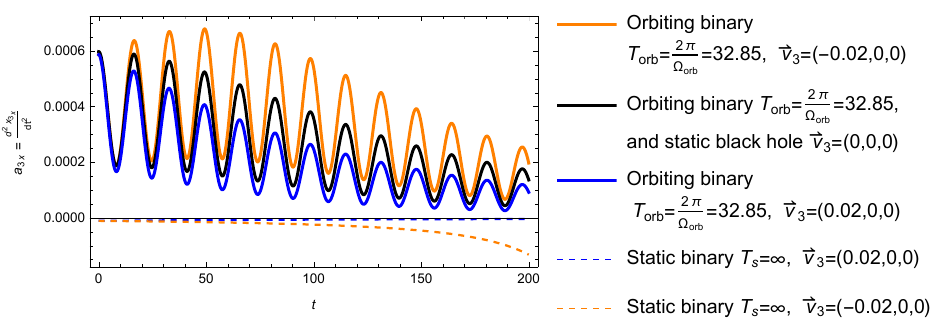}
 	\caption{Acceleration of the third black hole dropped from $\vec{x}(0)=(10,0,0)$, with a non-vanishing  initial velocity $\vec{v}_3(0)$ along the $x$-axis. We plot the $x$-component of the acceleration. The binary initially points towards the third hole. The phase of the waves present in the non-static case is the same for opposite velocities, which means the ripples originate in gravity and not in magnetism. We notice an attractive force if the binary is non-rotating (a consequence of relativistic mass increase irrespective of the direction of motion of the third black hole). The black curve is identical to the red curve in Figure
    \ref{fig:BlackHoleAcceleration}.}
\label{fig:BlackHoleAccelerationVelocityCompare}
\end{figure}	

\section{The 2-hole spacetime and its geodesics}\label{sec:Einstein-Maxwell_and_Geodesics}

Interestingly, as first pointed out in \citep{FeEa87,FeEa89}, the perturbative approach leads to an analytic form of a Majumdar-Papapetrou-like metric. The aim of this section is to explicitly evaluate the analytic prescription for the metric of two slowly moving extremally charged black holes moving steadily on a common circular orbit. For this solution, we verify it fulfills the Einstein-Maxwell equations. Subsequently, we shall look at the paths of various particles in the field of that orbiting pair.

\subsection{Explicit evaluation of the metric}

We now find an analytical form of the metric perturbation by solving equations \eqref{eq:P_Q_2} in terms of the two vector fields that we split into their divergence-free and curl-free parts $\vec{P}=\nabla \times \vec{\alpha} + \nabla \beta,$ and analogously with $\vec{Q}$. Since the curl-free parts of the $P, Q$ do not contribute to the action (see (14) in \cite{FeEa89}), we choose a gauge where they vanish identically, so that only the divergence free parts remains, yielding Laplace's equation for both, which we are able to solve in terms of integrals so that ultimately we find
 \begin{equation}
 \begin{aligned} \label{eq:rovnicePQ}
 \vec{P} (\vec{x}) &= \frac{3}{4\pi} \sum_{a=1}^{2} m_a 
 \int \psi^2(\vec{r'})
 \left[
 \vec{v}_a
 \frac{ \left(\vec{x}-\vec{r'} \right) \cdot \left( \vec{r'} -\vec{x}_a \right) }{|\vec{x}-\vec{r'} |^3 \ | \vec{r'} -\vec{x}_a |^3}
 -\left(\vec{v}_a \cdot \frac{ \left(\vec{x}-\vec{r'} \right) }{|\vec{x}-\vec{r'} |^3} \right) \frac{\left( \vec{r'} - \vec{x}_a \right)}{ | \vec{r'} -\vec{x}_a |^3}
 \right]
 \text{d} V',
 \\
 \vec{Q} (\vec{x}) &= \frac{1}{\pi} \sum_{a=1}^{2} m_a
 \int \psi^3(\vec{r'})
 \left[ 
 \vec{v}_a \frac{ \left(\vec{x}-\vec{r'} \right) \cdot \left( \vec{r'} -\vec{x}_a \right) }{|\vec{x}-\vec{r'} |^3 \ | \vec{r'} -\vec{x}_a |^3}
 -
 \left( \vec{v}_a \cdot \frac{ \left(\vec{x}-\vec{r'} \right) }{|\vec{x}-\vec{r'} |^3 }\right) \frac{\left( \vec{r'} - \vec{x}_a \right)}{| \vec{r'} -\vec{x}_a |^3}
 \right]
 \text{d} V',
 \end{aligned}
 \end{equation}
where $\psi (\vec{r}') = 1 + \sum_{a=1}^2 \frac{m_a}{|\vec{r}' - \vec{x}_a(t)|}$, $\vec{x}$ is the position in the three-space, $\vec{r}'$ is the integration variable over the 3-volume element $\text{d} V'$, and $a$ labels the black holes. Furthermore, \eqref{eq:rovnicePQ} can be rewritten in a more compact form as
 \begin{equation}
 \begin{aligned}
 \vec{P} (\vec{x}) &= \frac{3}{4\pi} \sum_{a=1}^{2} m_a 
 \!\! \int \!\! \psi^2(\vec{r'}) \frac{ \left(\vec{x}-\vec{r'} \right) \times \left( \vec{v}_a \times \left( \vec{r'} -\vec{x}_a \right)  \right)}{|\vec{x}-\vec{r'} |^3 | \vec{r'} -\vec{x}_a |^3}
 \text{d} V', 
 %
 \\ \vec{Q} (\vec{x})
 &= \frac{1}{\pi} \sum_{a=1}^{2} m_a 
 \!\! \int \!\! \psi^3(\vec{r'})
 \frac{ \left(\vec{x}-\vec{r'} \right) \times \left( \vec{v}_a \times \left( \vec{r'} -\vec{x}_a \right) \right) }{|\vec{x}-\vec{r'} |^3 | \vec{r'} -\vec{x}_a |^3}
 \text{d} V' . 
 \end{aligned}
 \end{equation}
We again split the integration domain as in Section \ref{sec:L3B}, keeping in mind that the domain of validity of the resulting metric is $|\vec{x}| \gg |\vec{r}_{21}|$.
Going through a rather long and tedious procedure and substituting the circular motion of the binary from the two-body section, we observe that the perturbation surprisingly vanishes up to $\mathcal{O}(v^2)$, with $\vec{N}=\vec{A}=0$. Hence, the circular motion now only manifests itself in the metric via the positions of the binary $\vec{x}_1(t),\vec{x}_2(t)$, which are explicit functions of time. The resulting metric of a slowly rotating, extremally charged binary thus is
 \begin{equation}
 \label{eq:PerturbedMetric}
 g_{\mu \nu}(t,\vec{x}) = \text{diag}(-\psi^{-2}, \psi^2,\psi^2,\psi^2), \quad
 A = \frac{1}{\psi} \text{d} t,
 \quad
 \psi (\vec{x},t) = 1 + \sum_{a=1}^2 \frac{m_a}{|\vec{x} - \vec{x}_a(t)|} .
 \end{equation}

The above metric and electromagnetic field satisfy Einstein equations with the right-hand side of the form
\begin{equation}
    T^{\mu\nu}_{\text{TOTAL}}=T^{\mu\nu}_{\text{DUST}}+T^{\mu\nu}_{\text{EM}} = \rho U^{\mu}U^{\nu} + \frac{1}{4 \pi} \left( F^{\mu \alpha} F^{\nu}_{\ \alpha} - \frac14 g^{\mu\nu} F^{\alpha \beta}F_{\alpha \beta} \right),
\end{equation}
with only the diagonal terms non-vanishing:
\begin{equation}
    G^{\mu\nu}-8\pi T^{\mu\nu}_{\text{TOTAL}} = \left(\begin{array}{cccc}
3 \psi^2  \dot{\psi}^2 & 0  & 0 & 0	\\
0 & -\frac{3 \psi^2  \dot{\psi}^2 + 2 \psi \ddot{\psi}}{\psi^2} & 0 & 0	\\
0 & 0 & -\frac{3 \psi^2  \dot{\psi}^2 + 2 \psi \ddot{\psi}}{\psi^2}  & 0	\\
0 & 0 & 0 & -\frac{3 \psi^2  \dot{\psi}^2 + 2 \psi \ddot{\psi}}{\psi^2}
\end{array}\right)  \approx \mathcal{O} (v^2),
\end{equation}
since for the 2 mutually orbiting black holes we have $\dot{\psi} = \mathcal{O}(v) $ and $\ddot{\psi} = \mathcal{O}(v^2),$ all the terms on the right-hand are of the second order in velocities and the Einstein equations indeed hold.

Denoting $\dot{\psi}=\partial_t \psi$, the Maxwell equations with $F=\text{d}A$ require
\begin{equation}
\label{eq:MaxwellEquations}
    \begin{aligned}
        \nabla_i F^{it} &= -\frac{\Delta \psi}{\psi^2} = 4 \pi J^t = 4 \pi \rho U^t,
        \quad
        \nabla_i F^{ij} = - \left( \frac{1}{\psi} \right)_{\!\!,i} \Gamma^j_{ti} + \left( \frac{1}{\psi} \right)_{\!\!,j} \Gamma^i_{ti} = + \frac{\nabla \dot{\psi}}{\psi^2} = 4 \pi J^j = 4 \pi \rho U^j,
    \end{aligned}
\end{equation}
where $\rho$ is the charge density of the dust equal to its rest mass density, and $U$ is its 4-velocity. These relations are consistent with the 4-velocity
\begin{equation} \label{eq:rovniceU}
    U^{\mu} = (\psi,\vec{v}),
\end{equation}
as expected with small $v^i$, since this keeps the time-like normalization $g_{\mu\nu}u^{\mu}u^{\nu} = -1 + \mathcal{O}(v^2)$ (unlike the choice in \citep{FeEa87}). With our choice we also get the right Laplace equation (Gauss's law)  $\Delta \psi = - 4 \pi \rho \psi^3$ from \citep{FeEa87}.
\subsection{Test--particle trajectories}
In this section the Greek indices $\alpha, \beta$ stand for $(t,x,y,z)$, the Latin indices $i,j$ denote spatial components, while $t$ as an index always denotes the time component of the quantity (no summation, etc.). The dot (e.g., $\dot{x}$) denotes the derivative with respect to the particle's proper time $\tau$, while the derivative w.r.t. the coordinate time is denoted by an index ($\psi_{,t}$).

Having an explicit metric \eqref{eq:PerturbedMetric} at hand, we can now solve the electrogeodesic equation, which we obtain as the Euler-Lagrange equation induced by the Lagrangian
 $ \mathcal{L} = \frac12 g_{\mu \nu} \dot{x}^{\mu} \dot{x}^{\nu} + \kappa \dot{x}^{\mu} A_{\mu}, $
where $\kappa$ is charge-to-mass ratio of the studied test particle.
The electrogeodesic equation ensures that the four-velocity normalization is conserved along the trajectory with
$g_{\mu \nu}u^{\mu} u^{\nu} = -1$ for time-like particles. We find
 \begin{align}
  \alpha = t:  \qquad & 
  -\frac{1}{\psi^2} \delta_{t \beta}\ddot{x}^{\beta}
  +  \frac{2}{\psi^3} \delta_{t \beta} \dot{x}^{\beta} \left( \dot{x}^{\alpha} \psi_{,\alpha} \right)  - \frac{\kappa}{\psi^2} \left( \dot{x}^{\alpha} \psi _{,\alpha} \right)
  =   \psi_{,t} \left (\frac{1}{\psi^3}   \delta_{t \alpha } \delta_{t \beta } \dot{x}^{\alpha} \dot{x}^{\beta}+  \psi \   \delta_{ij} \dot{x}^i \dot{x}^j     - \kappa  \ \delta_{t \beta} \dot{x}^{\beta} \frac{1}{\psi^2}    \right), \nonumber\\
   \alpha = i:   \qquad & \psi^2 \delta_{ij} \ddot{x}^j
   + 2 \psi \ \delta_{ij} \dot{x}^j \left( \dot{x}^{\alpha} \psi_{,\alpha}  \right) \nonumber
   =  \psi_{,i} \left (\frac{1}{\psi^3}   \delta_{t \alpha } \delta_{t \beta } \dot{x}^{\alpha} \dot{x}^{\beta}  +  \psi\    \delta_{ij} \dot{x}^i \dot{x}^j      - \kappa \ \delta_{t \beta} \dot{x}^{\beta} \frac{1}{\psi^2}    \right).
  \end{align}
The simplest possible solution is a static extremally charged particle with $\kappa=1$, which needs to satisfy
\begin{equation}
    -\frac{1}{\psi^2} \ddot{t} + \frac{\dot{t}^2}{\psi^3} \psi_{, t} = 0,
\end{equation}
yielding a simple relation $\dot{t} = \kappa \psi$. It is of interest that if we drop such a particle from rest, it will remain static indefinitely since all terms in the Taylor expansion of the spatial part of four-velocity as a function of the particle's proper time are proportional to $(\kappa-1)$. This is counterintuitive since the test particle might be affected by the motion of the central binary.

Choosing a non-extremal particle with a different $\kappa$, its motion is influenced already by the Coulomb-like interactions. For instance, a massive uncharged test particle with $\kappa=0$ is simply pulled towards the black holes due to the gravitational attraction. For trajectories of particles with various specific charges, see Figure \ref{fig:GeodesicPosition}. The following Figure \ref{fig:GeodesicAcceleration} shows the acceleration of the test particle---more specifically the second derivative of its position with respect to its proper time---as a function of the coordinate time.

\begin{figure}[ht]
	\centering
	\includegraphics[width = 8cm]{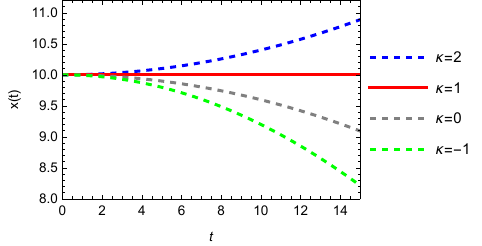}
 	\caption{Trajectories of initially static test  particles of various charge-to-mass ratios $\kappa$. Particles with an initial position $\vec{x}(0)=(10,10,10)$ are dropped from rest into the field of a black-hole pair orbiting steadily around the origin at a mutual distance $r_{21}=(\sqrt{3}-1)/2$. We plot the particles' coordinate $x$ as a function of the coordinate time $t$.
    An uncharged particle ($\kappa = 0$) and a particle charged oppositely than the central binary ($\kappa = -1$) are attracted towards the center, whilst a particle charged twice as much as the central pair ($\kappa=2$) is rather strongly repelled. An extremally charged particle  ($\kappa=1$) remains still, as per our analytical result. The Coulombic interaction dominates the wiggly effects seen in the following two plots.}
\label{fig:GeodesicPosition}
\end{figure}	

\begin{figure}[ht]
	\centering
	\includegraphics[width = 14cm]{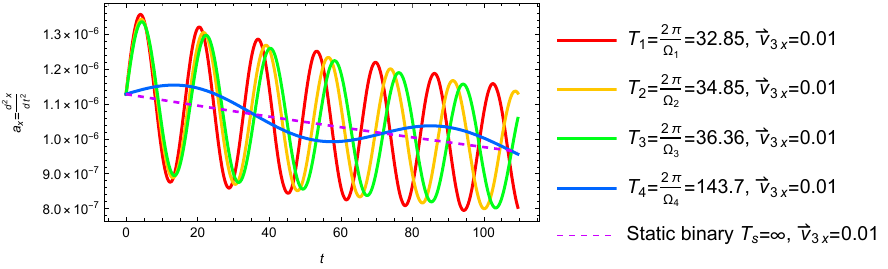}
 	\caption{ Acceleration  of an extremally charged test particle in the field of a black-hole pair orbiting with various periods $T_i$ (the second derivative of the test particle's position with respect to the coordinate time). The test particle is dropped from $\vec{x}(0)=(10,0,0)$ with the velocity $\vec{v}(0)=(0.01,0,0)$. We plot the $x$-component of the acceleration. The binary is initially pointing towards the test particle, which is heading away from the origin.}
\label{fig:GeodesicAcceleration}
\end{figure}	

\begin{figure}[ht]
	\centering
	\includegraphics[width = 14cm]{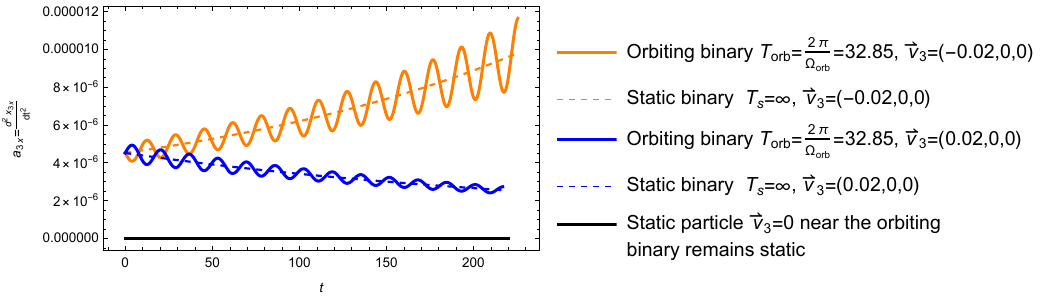}
 	\caption{ Acceleration  of an extremally charged test particle  in the field of a black-hole pair, either orbiting or static, and pointing towards the test particle initially. We compare two opposite initial velocities of the particle, see the legend. A positive velocity means the particle is heading away from the binary. We choose $\vec{x}(0)=(10,0,0)$; we plot the $x$-component of the acceleration.
    The phase of the ripples due to rotation is exactly opposite for opposite velocities.
    This phase is also shifted by a quarter period with respect to that in Figure \ref{fig:BlackHoleAccelerationVelocityCompare} for black hole with the same conditions.
    We also observe a repulsive force, resulting in a positive acceleration  for all the cases.
 }
\label{fig:GeodesicAccelerationVelocityCange}
\end{figure}	


\section{Differences between black-hole and test-particle trajectories}
When we compare the motion of an extremally charged test particle to that of a third black hole (which is also extremally charged), we discover an apparent self-force effect: dropping them both from rest in the field of the central black-hole binary, the test particle remains static while the black hole is repelled from the binary. This effect is manifest in Figure \ref{fig:BlackHoleAcceleration} where we present the motion of the third black hole which is initially static---and we know that a static particle stands still under the the same conditions as seen in Figure \ref{fig:GeodesicPosition}.
Figure \ref{fig:BlackHoleAccelerationVelocityCompare} shows that the self-force dominates the effect of relativistic mass increase, which would simply pull the hole towards the center, resulting in a negative acceleration.
The behavior observed in the test-particle case shown in Figure \ref{fig:GeodesicAccelerationVelocityCange} reveals a repulsive force as well, but it is two orders of magnitude smaller than for three black holes. As expected, when a particle approaches the central black-hole pair, it gets pushed out ever more strongly, and vice versa.

Comparing Figure \ref{fig:BlackHoleAccelerationVelocityCompare} with Figure \ref{fig:GeodesicAccelerationVelocityCange}, we observe that the acceleration of the black hole is two orders of magnitude greater than that of the test particle and the resulting curves are shifted by a quarter of the orbital period. The effects seen with the test particle are probably swamped by the back reaction of the third black hole. Note also the crucial difference between the orbiting and static cases in Figure \ref{fig:BlackHoleAccelerationVelocityCompare} describing the behavior of a third black hole, as opposed to Figure \ref{fig:GeodesicAccelerationVelocityCange} where the non-staticity of the central pair only causes small fluctuations around the static case.

\section{Conclusions}
In this paper, we extended the slow-motion perturbation of the Majumdar-Papapetrou solution from the case of a pair of extremally charged black holes to a triplet. We succeeded in finding the trajectory of a lesser companion coupled to a heavy central black-hole binary and compared it to the path of a charged test particle. The perturbed Majumdar-Papapetrou solution dictates that both an extremally charged test particle and a small third black hole are allowed to stand still if the central black-hole binary is static. Allowing the binary to orbit gives rise to a repulsive self-force acting upon the third black hole due to its contribution to the gravitational field, while the extremally charged test particle of negligible mass can still remain at rest even in this non static field.
Moreover, as the third hole is given any non-zero velocity, the relativistic effect of the increase in mass is noticed. Both the self-force and the mass increase are relativistic effects, which cannot be observed in Newtonian physics. Albeit we adopted some simplifying assumptions in our approach that allowed us to evaluate the required integrals, we still obtained results that conform to our intuition in some of the test-particle cases while revealing some new aspects of the restricted general relativistic 3-body problem, such as the repulsive force or the self-force acting upon the black-hole companion of the central binary.
\begin{acknowledgments}
E.K. was supported by the grant No. UNCE/24/SCI/016, M.Z. by the grant GACR 22-14791S. The symbolic and numerical calculations in this work were performed using Wolfram Mathematica \cite{WolframOne} and the xAct package \cite{xActWebsite}.
\end{acknowledgments}
%


\begin{thebibliography}{20}
\providecommand{\natexlab}[1]{#1}
\providecommand{\url}[1]{\texttt{#1}}
\expandafter\ifx\csname urlstyle\endcsname\relax
  \providecommand{\doi}[1]{doi: #1}\else
  \providecommand{\doi}{doi: \begingroup \urlstyle{rm}\Url}\fi

\bibitem[Shariat et~al.(2025)Shariat, El-Badry, and Naoz]{Shariat+El-Badry+Naoz}
Cheyanne Shariat, Kareem El-Badry, and Smadar Naoz.
\newblock {10,000 Resolved Triples from Gaia: Empirical Constraints on Triple Star Populations}.
\newblock \emph{Publications of the Astronomical Society of the Pacific}, 137\penalty0 (9):\penalty0 094201, sep 2025.
\newblock \doi{10.1088/1538-3873/adfb30}.
\newblock URL \url{https://doi.org/10.1088/1538-3873/adfb30}.

\bibitem[Matson et~al.(2018)Matson, Howell, Horch, and Everett]{Matson+Howell+Horch+Everett}
Rachel~A. Matson, Steve~B. Howell, Elliott~P. Horch, and Mark~E. Everett.
\newblock {Stellar Companions of Exoplanet Host Stars in K2}.
\newblock \emph{The Astronomical Journal}, 156\penalty0 (1):\penalty0 31, jun 2018.
\newblock \doi{10.3847/1538-3881/aac778}.
\newblock URL \url{https://doi.org/10.3847/1538-3881/aac778}.

\bibitem[Ledvinka et~al.(2008)Ledvinka, Sch\"afer, and Bi\v{c}\'ak]{Ledvinka+Schafer+Bicak}
Tom\'{a}\v{s} Ledvinka, Gerhard Sch\"afer, and Ji\v{r}\'{\i} Bi\v{c}\'ak.
\newblock {Relativistic Closed-Form Hamiltonian for Many-Body Gravitating Systems in the Post-Minkowskian Approximation}.
\newblock \emph{Phys. Rev. Lett.}, 100:\penalty0 251101, Jun 2008.
\newblock \doi{10.1103/PhysRevLett.100.251101}.
\newblock URL \url{https://link.aps.org/doi/10.1103/PhysRevLett.100.251101}.

\bibitem[Yamada and Asada(2010)]{Yamada+Asada}
Kei Yamada and Hideki Asada.
\newblock {Collinear solution to the general relativistic three-body problem}.
\newblock \emph{Phys. Rev. D}, 82:\penalty0 104019, Nov 2010.
\newblock \doi{10.1103/PhysRevD.82.104019}.
\newblock URL \url{https://link.aps.org/doi/10.1103/PhysRevD.82.104019}.

\bibitem[Imai et~al.(2007)Imai, Chiba, and Asada]{Imai+Chiba+Asada}
Tatsunori Imai, Takamasa Chiba, and Hideki Asada.
\newblock {Choreographic Solution to the General-Relativistic Three-Body Problem}.
\newblock \emph{Phys. Rev. Lett.}, 98:\penalty0 201102, May 2007.
\newblock \doi{10.1103/PhysRevLett.98.201102}.
\newblock URL \url{https://link.aps.org/doi/10.1103/PhysRevLett.98.201102}.

\bibitem[Lousto and Nakano(2008)]{Lousto+Nakano}
Carlos~O. Lousto and Hiroyuki Nakano.
\newblock {Three-body equations of motion in successive post-Newtonian approximations}.
\newblock \emph{Classical and Quantum Gravity}, 25\penalty0 (19):\penalty0 195019, sep 2008.
\newblock \doi{10.1088/0264-9381/25/19/195019}.
\newblock URL \url{https://doi.org/10.1088/0264-9381/25/19/195019}.

\bibitem[Huang and Wu(2014)]{Huang+Wu}
Guoqing Huang and Xin Wu.
\newblock {Dynamics of the post-Newtonian circular restricted three-body problem with compact objects}.
\newblock \emph{Phys. Rev. D}, 89:\penalty0 124034, Jun 2014.
\newblock \doi{10.1103/PhysRevD.89.124034}.
\newblock URL \url{https://link.aps.org/doi/10.1103/PhysRevD.89.124034}.

\bibitem[Strong and Crescimanno(2020)]{Strong+Crescimanno}
Martin~D. Strong and Michael Crescimanno.
\newblock {Lagrange point stability for a rotating host mass binary}.
\newblock \emph{Phys. Rev. D}, 102:\penalty0 024052, Jul 2020.
\newblock \doi{10.1103/PhysRevD.102.024052}.
\newblock URL \url{https://link.aps.org/doi/10.1103/PhysRevD.102.024052}.

\bibitem[{Ferrell} and {Eardley}(1987)]{FeEa87}
Robert~C. {Ferrell} and Douglas~M. {Eardley}.
\newblock {Slow-motion scattering and coalescence of maximally charged black holes}.
\newblock \emph{Phys. Rev. Lett.}, 59:\penalty0 1617--1620, Oct 1987.
\newblock URL \url{https://link.aps.org/doi/10.1103/PhysRevLett.59.1617}.

\bibitem[{{Ferrell}} and {{Eardley}}(1989)]{FeEa89}
Robert~C. {{Ferrell}} and Douglas~M. {{Eardley}}.
\newblock \emph{{Slowly moving maximally charged black holes in ``Frontiers in Numerical Relativity''}}, pages 27--42.
\newblock {C.~R. {Evans}, L.~S. {Finn}, and D.~W. {Hobill},} (eds.) Cambridge University Press, 1989.

\bibitem[{Camps} et~al.(2017){Camps}, {Hadar}, and {Manton}]{CHM}
Joan {Camps}, Shahar {Hadar}, and Nicholas~S. {Manton}.
\newblock {Exact gravitational wave signatures from colliding extreme black holes}.
\newblock \emph{Phys. Rev. Lett. D}, 96, 2017.
\newblock URL \url{https://doi.org/10.1103/PhysRevD.96.061501}.

\bibitem[{McCarthy} et~al.(2018){McCarthy}, {Kubiz{\v n}{\'a}k}, and {Mann}]{FionaDaKu}
Fiona {McCarthy}, David {Kubiz{\v n}{\'a}k}, and Robert~B. {Mann}.
\newblock Dilatonic imprints on exact gravitational wave signatures.
\newblock \emph{Physical Review D}, 97\penalty0 (10), May 2018.
\newblock URL \url{https://doi.org/10.1103/PhysRevD.97.104025}.

\bibitem[{Majumdar}(1947)]{Ma}
Sudhansu~D. {Majumdar}.
\newblock {A Class of Exact Solutions of Einstein's Field Equations}.
\newblock \emph{{Physical Review}}, 72\penalty0 (5):\penalty0 390--398, 1947.
\newblock ISSN 0031-899X.
\newblock URL \url{https://link.aps.org/doi/10.1103/PhysRev.72.390}.

\bibitem[{Papapetrou}(1945)]{Pa}
Achilleas~N. {Papapetrou}.
\newblock {A Static Solution of the Equations of the Gravitational Field for an Arbitary Charge-Distribution}.
\newblock \emph{Proceedings of the Royal Irish Academy. Section A: Mathematical and Physical Sciences}, 51:\penalty0 191--204, 1945.
\newblock ISSN 00358975.
\newblock URL \url{http://www.jstor.org/stable/20488481}.

\bibitem[{Hartle} and {Hawking}(1972)]{HaHa}
James~B. {Hartle} and Stephen~W. {Hawking}.
\newblock {Solutions of the Einstein-Maxwell Equations with Many Black Holes}.
\newblock \emph{{Commun. Math. Phys.}}, 26:\penalty0 87--101, 01 1972.
\newblock URL \url{https://doi.org/10.1007/BF01645696}.

\bibitem[{Gibbons} and {Ruback}(1986)]{GiRu86}
Gary~W. {Gibbons} and Peter~J. {Ruback}.
\newblock {Motion of Extreme Reissner-Nordstrom Black Holes in the Low-Velocity Limit}.
\newblock \emph{Phys. Rev. Lett.}, 57:\penalty0 1492--1495, Sep 1986.
\newblock URL \url{https://doi.org/10.1103/PhysRevLett.57.1492}.

\bibitem[{{Klime{\v s}ov{\'a}}}(2023)]{diplomka}
Eli\v{s}ka {{Klime{\v s}ov{\'a}}}.
\newblock {\textit{Multi-black-hole gravitational field.}{ Master thesis, Charles University Prague, Institute of Theoretical Physics}}, 2023.
\newblock URL \url{https://dspace.cuni.cz/handle/20.500.11956/183832}.

\bibitem[Inc.()]{WolframOne}
Wolfram~Research{,} Inc.
\newblock {Wolfram|One, Version 15.0}.
\newblock URL \url{https://www.wolfram.com/wolfram-one}.
\newblock Champaign, IL, 2026.

\bibitem[Mart{\'i}n-Garc{\'i}a()]{xActWebsite}
Jos{\'e}~M. Mart{\'i}n-Garc{\'i}a.
\newblock {xAct: Efficient tensor computer algebra for the Wolfram Language}.
\newblock \url{http://www.xact.es/}.
\newblock Accessed: 2026-07-31.

\end{thebibliography}
\end{document}